\documentclass[runningheads]{llncs}
\usepackage[T1]{fontenc}
\usepackage{graphicx}
\usepackage{siunitx}
\usepackage{tabularray}
\usepackage{booktabs, tabularx, array, enumitem}
\usepackage{subfigure}
\usepackage{multirow}
\usepackage[normalem]{ulem}
\usepackage{hhline}

\usepackage{amssymb}
  \usepackage{graphicx}
  \usepackage{amsmath}
  \usepackage{siunitx}
  \usepackage{bbding}
  \usepackage{caption} 

\newcolumntype{L}[1]{>{\raggedright\arraybackslash}p{#1}} 
\newcolumntype{Y}{>{\raggedright\arraybackslash}X}

\begin{document}

\title{Design Implications for Student and Educator Needs in AI-Supported Programming Learning Tools}
\titlerunning{Design Implications for AI-Supported Programming Learning Tools}
%

\author{
Boxuan Ma\inst{1}
\and
Yinjie Xie\inst{1}
\and
Huiyong Li\inst{1}
\and
Gen Li\inst{1}
\and
Li Chen\inst{2}
\and \\
Atsushi Shimada\inst{1}
\and
Shin'ichi Konomi\inst{1}}
\authorrunning{Ma et al.}

\institute{Kyushu University, Fukuoka, Japan  \\
\and
Osaka Kyoiku University, Osaka, Japan
}
%

%
\maketitle        

\begin{abstract}
AI-powered coding assistants can support students in programming courses by providing on-demand explanations and debugging help. However, existing research often focuses on individual tools, leaving a gap in evidence-based design recommendations that reflect both educator and student perspectives in education settings. To ground the design of learning-oriented AI coding assistants for both sides' needs, we conducted parallel surveys of educators (N=50) and students (N=90) to compare preferences about (i) how students should request help, (ii) how AI should respond, and (iii) who should control. Our results show that educators generally favored indirect scaffolding that preserves students’ reasoning, whereas students were more likely to prefer direct, actionable help. Educators further highlighted the need for course-aligned constraints and instructor-facing oversight, while students emphasized timely support and clarity when stuck. Based on these findings, we discuss the interaction-focused design space and derive design implications for learning-oriented AI coding assistants, highlighting scaffolding and control mechanisms that balance students’ agency with instructional constraints.

\end{abstract}

\section{Introduction}

AI-driven coding assistance is becoming common in programming education, motivated by the difficulty many beginners face when learning to program \cite{li2025examining}. Alongside established supports such as intelligent tutors, programming games, and feedback-based learning platforms \cite{crow2018intelligent}, AI tools offer scalable, on-demand help that can adapt to students’ immediate questions and contexts \cite{ma2024enhancing}.

However, concerns about student over-reliance on AI, together with evidence that many students struggle to write effective prompts and may therefore receive unhelpful feedback \cite{ma2024enhancing,li2025coderunner}, have motivated researchers to develop learning-oriented tools that better align AI assistance with pedagogical goals. Representative examples include Coding Steps \cite{kazemitabaar2023studying}, CodeHelp \cite{liffiton2023codehelp}, Promptly \cite{denny2024prompt}, and CodeAid \cite{kazemitabaar2024codeaid}. Together, these systems explore complementary scaffolding strategies that incorporate stepwise hints \cite{kazemitabaar2023studying}, prompt-construction practice for code generation \cite{denny2024prompt}, and constraints on direct answers \cite{liffiton2023codehelp,kazemitabaar2024codeaid} to support students’ progress while reducing opportunities for answer copying.

Despite this progress, much of the literature has centered on building and evaluating individual systems, with comparatively less emphasis on deriving broader, transferable design guidance grounded in the expectations of both students and educators \cite{wu2025learner}. When design implications are discussed, they are often incomplete—either reflecting the needs of a single stakeholder group \cite{kazemitabaar2024codeaid,ma2025scaffolding} or addressing only a narrow set of design decisions \cite{wu2025learner}. Consequently, we still lack systematic, stakeholder-grounded comparisons of how educators and students prioritize key design choices for AI programming help, and where their preferences converge or diverge.


To address this gap and characterize what effective AI programming help in educational settings should look like, we investigate four research questions that arise when deploying AI coding assistants in programming courses:
\begin{itemize}[label=\textbullet]
\item \textbf{RQ1 (AI Policy and Perceptions):} What AI-use policies are considered appropriate in course settings, and how do stakeholders perceive the learning value of AI assistance?
\item \textbf{RQ2 (Learner Input and Context):} How should students formulate requests, and should systems automatically incorporate course-specific context or require students to provide it?
\item \textbf{RQ3 (AI Responses and Scaffolding):} Should assistants provide direct solutions or indirect scaffolding (e.g., hints, step-by-step plans), and what should be the maximum level of scaffolding permitted?
\item \textbf{RQ4 (Control and Responsibility):} Who should control the level and type of scaffolding in a course setting—the instructor, the student, or the system?
\end{itemize}


To answer these questions, we conducted parallel surveys with two stakeholder groups: educators with experience teaching programming (N=50) and students with programming learning experience (N=90). We compare where educator and student views converge or diverge and derive design implications for learning-oriented AI coding assistants. 

\begin{figure}[t]
\centering
\includegraphics[scale=0.27]{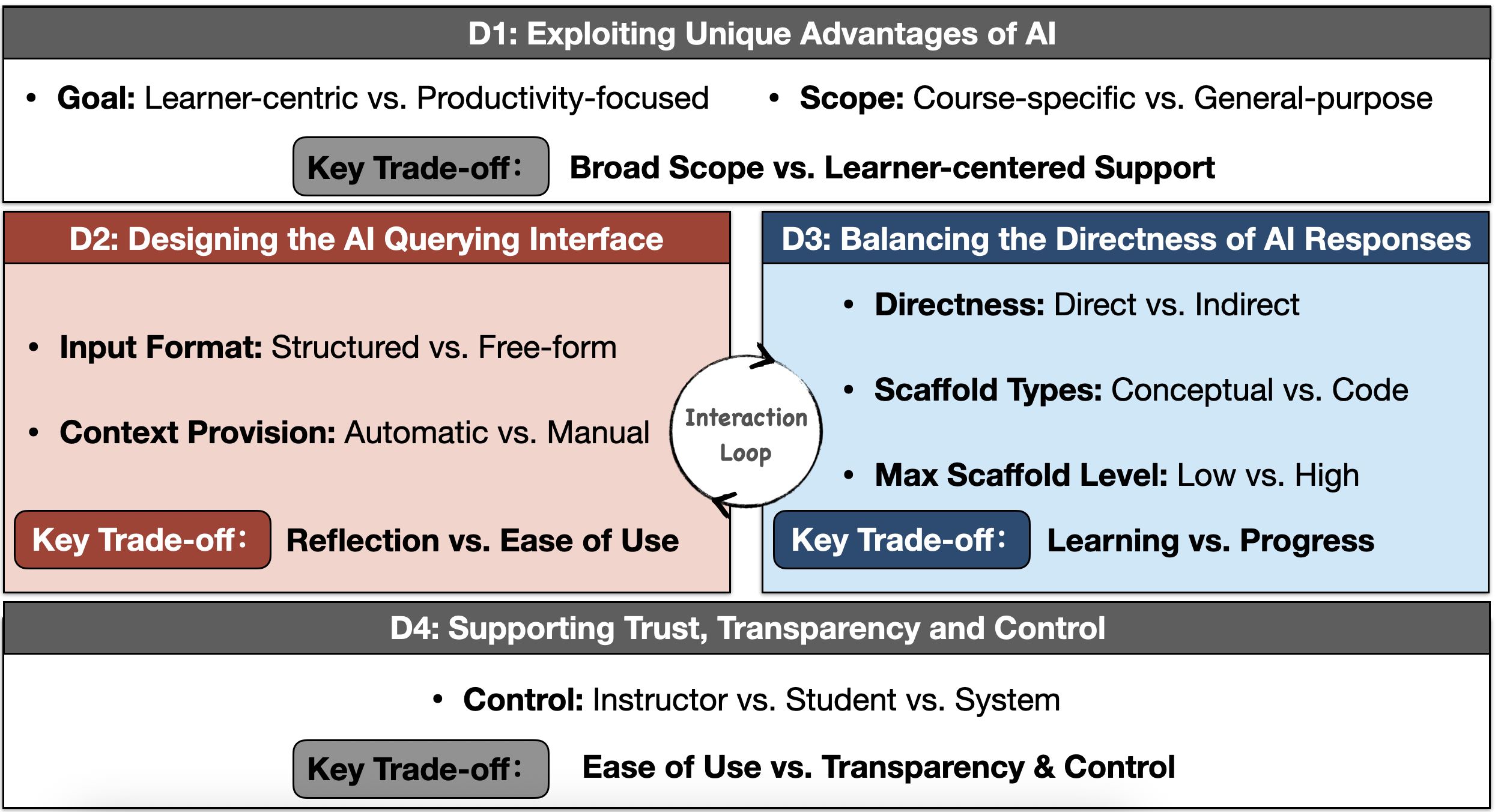}
\caption{Design considerations and trade-offs within the design space of AI-powered assistants for educational settings. Each consideration is based on a key stage in students’ help-seeking process.}
\label{design}
\end{figure}

\section{Design Space for AI Coding Assistants}

The four research questions above reflect recurring decisions educators and students face when AI is used during programming practice: when to rely on AI, what to provide as input, how the assistant should respond, and who should set and control boundaries. To connect these decisions into a coherent interaction loop and to ground our survey measures, we synthesize an interaction-focused design space with four considerations (Figure~\ref{design}). This framing is inspired by prior discussions of pedagogical AI coding assistants \cite{kazemitabaar2024codeaid,ma2025scaffolding}, but we adapt the dimensions to highlight classroom constraints and to make the underlying trade-offs explicit, providing an organizing lens for interpreting where educator and student preferences converge or diverge.




\subsection{Exploiting Unique Advantages of AI (D1):}

An initial stage of the help-seeking process is deciding what learning resource to use within the broader learning ecosystem available to them. This leads to the first major design consideration of an AI coding assistant: determining the role, scope, and unique advantages of future educational AI assistants in relation to other educational resources, like TA office hours, discussion boards, textbooks, etc \cite{kazemitabaar2024codeaid}. 

We need to think about the \textbf{Goal} and \textbf{Scope} of the AI assistant. An assistant can be designed primarily to maximize throughput (e.g., quickly producing working code) or to promote learning processes (e.g., prompting reflection, explaining reasoning, encouraging verification). A general-purpose assistant supports a wide range of questions but may drift from course expectations. A course-specific assistant can leverage assignment requirements, learning objectives, and instructor-intended approaches, potentially improving relevance and reducing misleading guidance. More broadly, these choices are closely tied to how instructors and students perceive the role of AI in learning, and to the course’s AI-use policy that translates those expectations into practice.

\subsection{Designing the AI Querying Interface (D2)}
Once students decide to use an assistant, the next question is how they should express their needs and what context should be available at query time. We focus on two design dimensions to balance user-friendliness with meta-cognitive engagement.

\textbf{Input format (structured vs.\ free-form).}
Whether the assistant allows students to ask questions in a free-form manner, similar to interaction with popular chat-based AI models, or requires input in a specific structure and format, is an important design consideration. Free-form prompts minimize friction, which is simple and easy to use. However, many students, especially novices, struggle to formulate effective prompts: they may ask overly vague questions, resulting in generic or misleading outputs. In contrast, structured prompts (e.g., templates or checklists) can elicit higher-quality requests by nudging students to specify goals, prior attempts, thereby externalizing metacognitive work and improving the relevance of AI feedback. The trade-off is that structured input can feel burdensome or restrictive: it increases interaction overhead, and can reduce flexibility when students want to ask quick, informal questions. 

\textbf{Context provision (automatic vs.\ manual).}
Whether the assistant automatically gathers context (like a plugin in an IDE) or requires manual input from the user is another important design consideration. An assistant may automatically gather relevant course context (e.g., learning context, assignment spec, rubric, IDE signals) or require students to input context manually. Automatic context reduces missing information and effort, but may limit opportunities for students to formulate problems and think critically. Manual context increases transparency and student agency but raises interaction cost and failure risk when students omit key details.

\subsection{Balancing the Directness of AI Responses (D3)}
After a prompt is sent, the AI’s response shapes both progress and learning. We frame this as balancing how direct the help is, what kind of scaffolding it uses, and how far the system is allowed to go.

\textbf{Directness (direct vs.\ indirect).} 
Direct responses prioritize rapid progress and can reduce short-term frustration by immediately unblocking students. However, they also increase the risk of copy\&paste behaviors and "answer outsourcing," where students accept or submit AI-generated code with minimal comprehension. This can shorten productive struggle, encourage shallow learning, and weaken opportunities tailored to their zone of proximal development. In contrast, indirect responses are intended to preserve students’ reasoning processes by prompting them to diagnose, revise, and implement solutions on their own. Such scaffolding can support self-repair and metacognitive engagement, but it comes with usability costs: indirect responses can risk discouraging students if they feel overwhelmed, unsupported, and not making progress. Prior work suggests that students strongly value timely, actionable guidance and may disengage when it feels indirect or slow. If guidance is too slow to translate into actionable code changes, students may perceive the tool as unhelpful, and revert to more direct tools or requests for complete answers~\cite{ma2025generative}. This introduces an important trade-off: finding the right balance between sufficient scaffolding for critical learning and minimizing frustration, while also considering the degree of autonomy students should have in personalizing their level of scaffolding.

\textbf{Scaffold types and max scaffold level.}
Even within indirect help, scaffolding can emphasize conceptual support (e.g., explaining underlying ideas, prompting reasoning, or providing simple hints) or code-oriented guidance (e.g., pseudo-code, partial code, or worked examples). Conceptual scaffolds can promote schema building and transfer by helping students articulate what they know, surface misconceptions, and reason about next steps, however, they may feel “too abstract” when students are stuck on concrete implementation details. By contrast, code-oriented scaffolds can be more immediately actionable and reduce cognitive load by making solution structure visible, but they also move closer to answer-giving. Importantly, these trade-offs are not one-size-fits-all. Students differ in prior knowledge, confidence, and help-seeking habits, and thus may prefer different scaffold types for the same task.

\subsection{Supporting Trust, Transparency and Control (D4)}
Finally, assistance in classrooms must be governable: stakeholders need clarity about what the assistant is allowed to do and who decides the level of help. A key challenge in AI-supported adaptive learning systems is determining how much decision-making power and responsibility should be given to the student versus the system \cite{imhof2020implementation}.

\textbf{Control (instructor vs.\ student vs.\ system).}
Previous work proposes two dimensions of learner-system control model: (1) the fading of scaffolding as they become more experienced (levels of help), and (2) managing their preferences for the different types of help available (types of help) \cite{wu2025learner}. Control may be instructor-driven, student-driven (self-selected), system-driven (automatic), or shared. If the system or instructor has full control, students might feel frustrated when the system’s intervention does not align with their expectations. Conversely, if students have full control, they may feel overwhelmed by the responsibility of choosing appropriate scaffolding—or strategically “game” the system by lowering support constraints to obtain near-direct answers.

Taken together, these design considerations provide a compact design space for interpreting where educator and student preferences converge or diverge, and for translating those patterns into actionable implications for AI coding assistants in education settings.

\section{Method}

\subsection{Survey Design and Measures}
We conducted parallel surveys with educators and students to understand their preferences for AI-based coding assistants in programming education. The survey was grounded in the design considerations in Figure~\ref{design}. To ensure classroom relevance and coverage, we iteratively developed the survey with four computing educators through multiple rounds of refinement, focusing on clarity, interpretability, and alignment with common teaching constraints. 

Both educator and student surveys included (i) background questions (e.g., teaching/learning experience, course types, programming languages), and (ii) preference questions aligned with the design space. For \textbf{AI policy and perceptions} (RQ1), participants reported their preferred AI-use policy in course settings and rated perceived learning benefits of AI assistance. For \textbf{Learner input and context} (RQ2), participants rated the perceived learning value of different student question types, indicated whether a structured input form versus free-form prompting better supports high-quality thinking, and whether course-specific context should be provided manually by students or gathered automatically by the system. For \textbf{AI responses and scaffolding} (RQ3), participants compared preferences for direct solutions versus indirect scaffolding, rated the helpfulness of different scaffolding types, and selected the maximum scaffolding level they would choose in a course context. For \textbf{control and responsibility} (RQ4), participants indicated who should control the scaffolding type and level. Both surveys included Likert-scale items and open-ended questions to capture preferences as well as their rationales. Detailed survey items are given in supplementary material\footnote{https://bit.ly/4kdk67M \label{appendix}}.

\subsection{Procedure and Recruitment}
We administered two parallel online surveys: one for educators and one for students. We distributed the survey link both via an online survey platform and through snowball sampling. Respondents were primarily based in the United States, with a smaller portion located in Asia, mainly Japan and China. Both surveys began with screening items to confirm eligibility, followed by the background and preference sections described above. Participants were informed of the study purpose and provided consent before proceeding. 

\subsection{Participants}

\subsubsection{Educator participants}
We surveyed 50 educators with experience teaching programming or computing-related courses. Teaching experience ranged from novice to highly experienced: 17 (34.0\%) reported 0--1 years, 13 (26.0\%) reported 1--3 years, 13 (26.0\%) reported 3--5 years, 5 (10.0\%) reported 5--10 years, and 2 (4.0\%) reported 10--20 years. Regarding teaching level, 33 (66.0\%) taught undergraduate courses only, 14 (28.0\%) taught graduate courses only, and 3 (6.0\%) taught at both levels. For course types, 29 reported teaching introductory programming and 22 reported teaching data science / machine learning (multiple selections allowed). Most educators used Python in their instruction (24), followed by Java (18) and C/C++ (11) (multiple selections allowed).

\subsubsection{Student participants}
A total of 90 students with experience in programming-related courses participated in the survey. In terms of academic level, 47 (52.2\%) were undergraduates, 38 (42.2\%) were graduate students, and 5 (5.6\%) selected other. Gender was approximately balanced (45 female, 50.0\%; 44 male, 48.9\%; 1 prefer not to say, 1.1\%). Regarding major, 21 (23.3\%) reported a computing-related field (e.g., CS/IT/DS), 20 (22.2\%) reported other STEM majors, 20 (22.2\%) reported non-STEM majors, and 29 (32.2\%) selected other. Programming experience skewed toward beginners: 50 (55.6\%) reported 0--6 months, 20 (22.2\%) reported 6--12 months, 8 (8.9\%) reported 1--2 years, and 12 (13.3\%) reported 2+ years.


\section{Results}

\subsection{AI-Policy and Perceptions}

Educators predominantly supported either “allowed with conditions” or “mostly open” policies for student AI use in programming courses : 55.1\% selected allowed with conditions, 26.5\% selected mostly open, and 18.4\% selected prohibited. In the open-ended explanations, conditional policies were typically justified as a way to preserve students’ reasoning while still leveraging AI as a support tool, but not as a substitute for producing final answers (e.g., “use it as support but not to make all the work at once”; “it cannot be your final answer”). Several educators also emphasized accountability mechanisms, such as requiring students to explain AI-produced code, or to disclose AI use when appropriate. A smaller set of educators adopted stricter stances (prohibited) primarily to ensure authentic learning and prevent shortcutting, while more open policies were often motivated by the view that students—especially at advanced levels—should learn to use AI responsibly and that access to AI support can reduce barriers and inequities. 

Students similarly leaned toward conditional allowance as the preferred policy for courses. Specifically, 55.6\% selected allowed with conditions, 20.0\% selecting mostly open, 16.7\% reporting not sure, and 7.8\% selecting prohibited. This distribution suggests that students generally accept AI assistance in coursework, but still recognize the importance of explicit rules distinguishing learning support from answer submission.

\begin{figure}[t]
\centering
\includegraphics[scale=0.29]{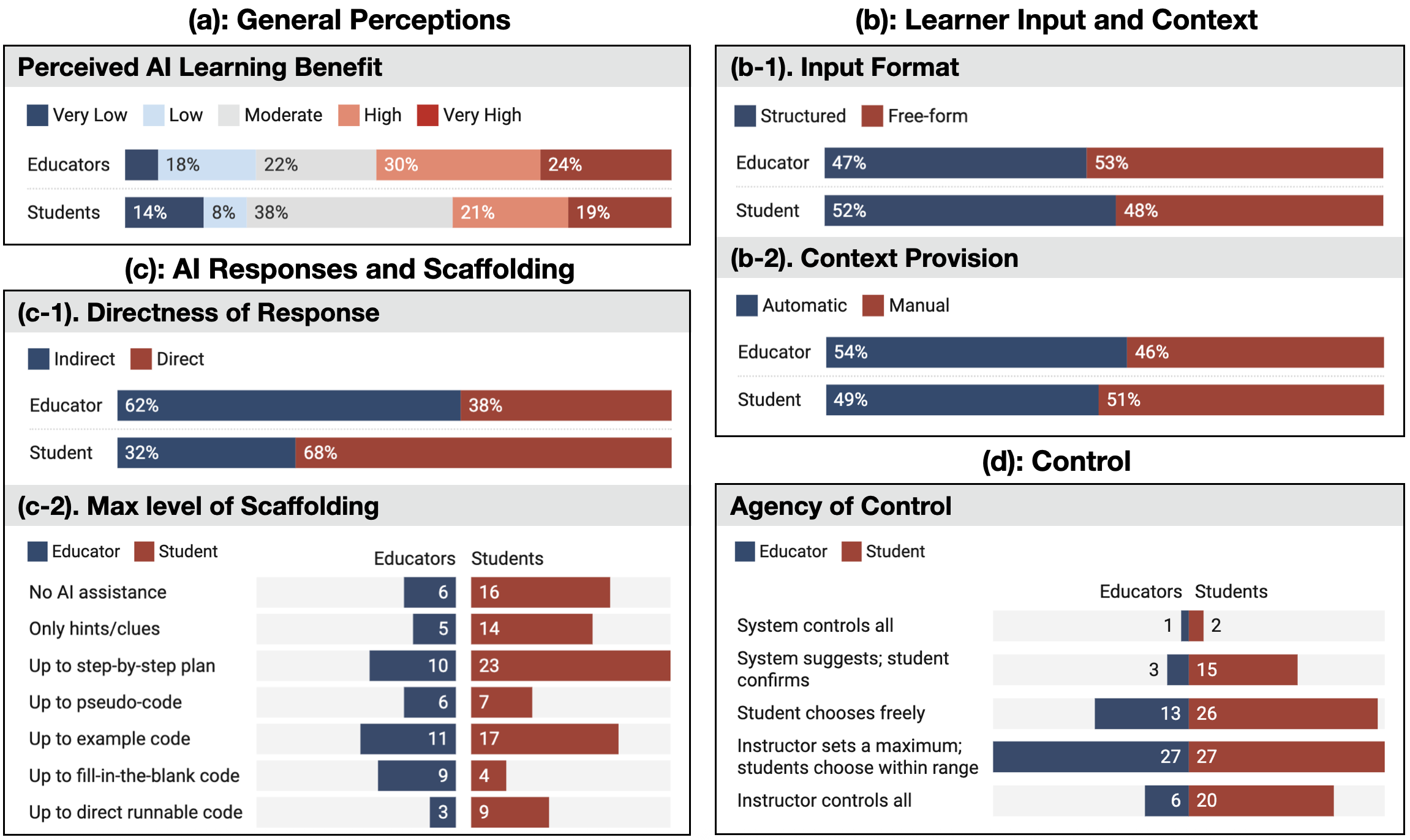}
\caption{Survey results overview: educator and student preferences for AI coding assistants for educational setting.}
\label{summary}
\end{figure}

In terms of general perceptions, educators’ were cautiously positive but mixed. As shown in Figure~\ref{summary}(a), 53.1\% agreed or strongly agreed, 22.4\% were neutral, and (24.5\%) disagreed or strongly disagreed. This pattern aligns with their qualitative comments: many acknowledged AI’s value for timely support and explanation, while simultaneously stressing the need for boundaries that discourage over-reliance and solution outsourcing. In contrast, students were more ambivalent than educators, 40.0\% agreed or strongly agreed, 37.8\% were neutral, and 22.2\% disagreed or strongly disagreed. 

\subsection{Learner Input and Context}

\subsubsection{Input Formats for AI Use.}

As shown in Figure~\ref{summary}(b-1), when asked which input format better promotes high-quality thinking when students use AI, educators were split but leaned slightly toward free-form input (53\%) over structured forms (47\%). Educators who favored structured formats emphasized that templates/checklists can scaffold students to articulate their understanding, avoid underspecified or misleading prompts, and thereby improve learning quality. One educator noted that ``Well-designed, structured input forms can scaffold better prompts and improve comprehension.'' Others highlighted that structured formats are particularly important for students with weak programming backgrounds, as they might otherwise resort to directly asking AI for answers without sufficient reasoning. In contrast, educators who preferred free-form inputs pointed to accessibility and student autonomy. They argued that free-form interactions encourage more frequent questioning and that “asking AI effectively is also an important skill.” One educator noted that because free-form input mirrors real-world tools such as ChatGPT, it can lower barriers and promote engagement, even if the questions vary in quality.

On the other hand, students showed a similar split, with a slight preference for structured input (52\%) over free-form input (48\%). Students who favored structured input likely value the clarity and guidance that reduces uncertainty about what to include (e.g., task goals, error message), which can make AI help feel more actionable. Meanwhile, students who preferred free-form interaction prioritized speed and convenience, especially when they were stuck and needed quick clarification, and perceived structured templates as additional overhead during problem-solving.

\subsubsection{Context Provision for AI Use.}

As shown in Figure~\ref{summary}(b-2), educators slightly preferred automatically gathering course-specific context (54\%) over requiring manual input from students (46\%). Those favoring automatic context provision viewed it as a way to improve response relevance and reduce the burden on students to supply missing details (e.g., assignment requirements, allowed libraries, or course conventions), which can be difficult for novices to specify precisely. At the same time, a sizable share favored manual input, noting that fully automated context gathering could reduce opportunities for student reflection and sense-making, and could also raise classroom-management concerns such as overreliance.

Students were nearly evenly divided but leaned slightly toward manual context provision (51\%) over automatic provision (49\%). This pattern suggests that many students may want to retain agency over what information is shared with the AI (e.g., what parts of their code, errors, or task descriptions are included), potentially due to privacy, control, or a desire to keep interactions lightweight. Conversely, students who favored automatic context prioritized convenience and faster, more accurate help without needing to curate detailed descriptions.

\begin{figure}[t]
\centering
\includegraphics[scale=0.35]{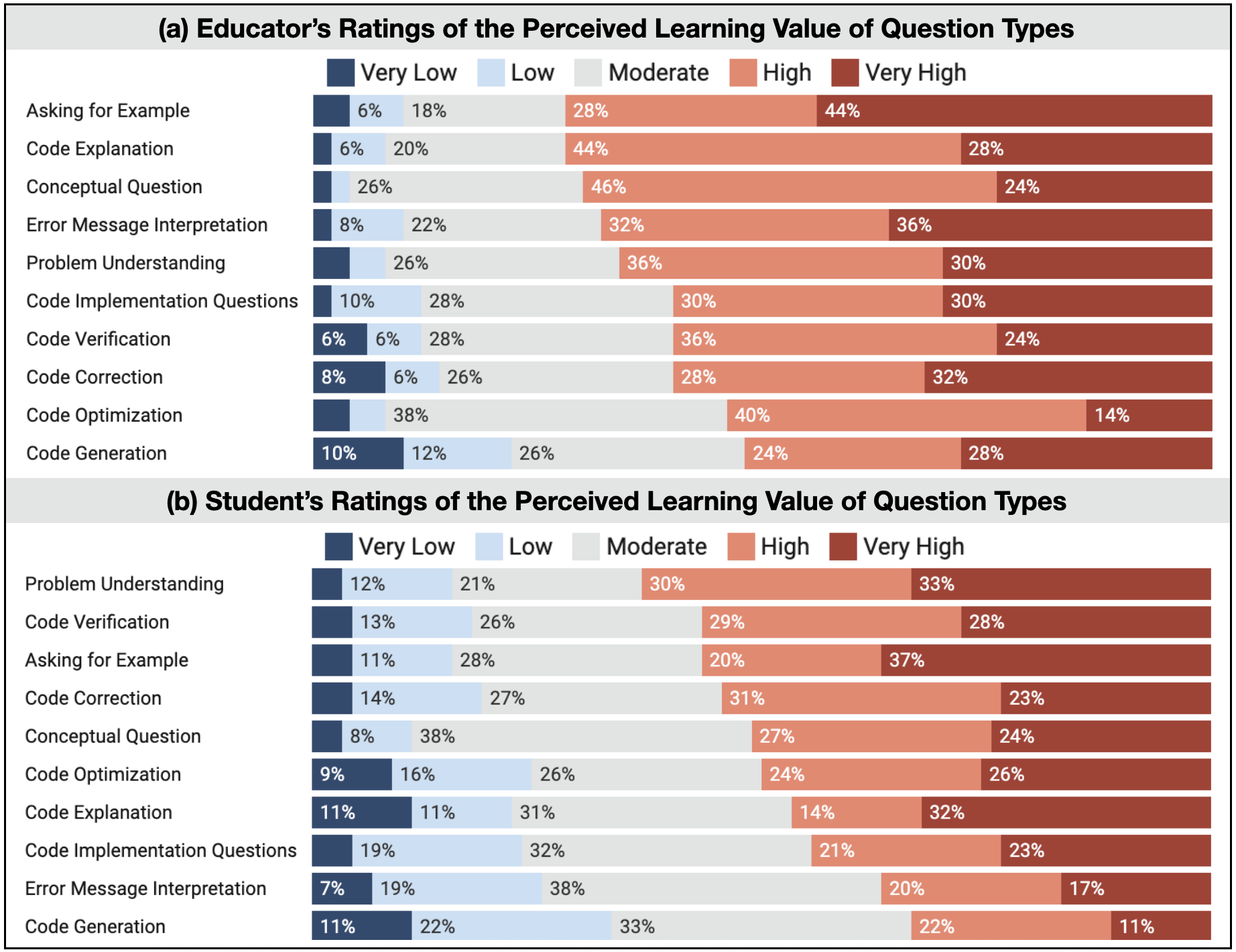}
\caption{(a) Educators’ and (b) students’ perceived learning value ratings by question type. Items are ordered by the sum of High and Very High responses.}
\label{survey1}
\end{figure}

\subsubsection{Perceived Learning Value of Question Types.}

Understanding which prompt types are perceived as learning-productive clarifies what an AI assistant should encourage through its input interface and how it should map requests to appropriate response scaffolds. We therefore report educator and student ratings separately to identify convergences and tensions that directly motivate our stakeholder-grounded design implications. The question-type taxonomy follows prior work coding authentic programming help-seeking and is grounded in a codebook derived from real student questions, full codes and definitions are provided in supplementary material\footref{appendix}. We report educator and student ratings separately to capture each group’s perceptions.

Figure~\ref{survey1}(a) shows educators' ratings of the perceived learning value of different student question types. Overall, educators tended to assign higher value to prompts that support sense-making and articulation than to prompts that directly outsource solution production. In particular, \emph{Asking for Example}, \emph{Code Explanation}, \emph{Conceptual Question} and \emph{Problem Understanding} were among the highest-rated categories. For debugging-related prompts, \emph{Error Message Interpretation} was generally valued, while \emph{Code Verification} and \emph{Code Correction} received more mixed evaluations. By contrast, \emph{Code Generation} was rated lowest overall, and \emph{Code Optimization} also attracted more reserved ratings than foundational learning activities. This suggests that educators prioritize prompts that keep students cognitively engaged, such as understanding why code works, diagnosing issues, and connecting new concepts to prior knowledge. 

Figure~\ref{survey1}(b) shows students’ ratings, which reflect a more progress and action oriented view of help-seeking.  Students rated \emph{Problem Understanding}, \emph{Code Verification}, \emph{Asking for Example}, and \emph{Code Correction} highest, indicating that many students value help that clarifies task requirements and provides concrete reference points, while still supporting checking and confirmation. In other words, rather than valuing prompts solely for conceptual elaboration, students appear to favor inputs that help them move forward when blocked, while retaining a sense of control through verification and confirmation. In addition, students were more negative about several code-oriented categories. Notably, \emph{Error Message Interpretation} received relatively lower ratings, and \emph{Code Implementation Questions} were also evaluated less favorably than educators did. \emph{Code Generation} was rated lowest by students as well, showing that even students do not necessarily view direct code output as the most educationally valuable form of help.

\subsection{AI Responses and Scaffolding}

\subsubsection{Preferences for Response Directness}

As shown in Figure~\ref{summary}(c-1), educators showed a clear preference for indirect assistance: 62\% favored indirect scaffolding over direct responses (38\%). This pattern aligns with the common instructional rationale that maintaining a degree of productive struggle can protect students’ reasoning and reduce the likelihood of answer-copying. Educators’ comments also suggested that indirect formats are easier to align with course policies, because they support learning while still keeping students responsible for producing the final solution.

Students, by contrast, more often preferred direct responses: 68\% favored direct help, whereas 32\% preferred indirect scaffolding. This preference reflects students’ immediate need for forward progress when stuck (e.g., resolving errors quickly or meeting deadlines), as well as the fact that many students experience high cognitive load during implementation and debugging. 

\begin{figure}[t]
\centering
\includegraphics[scale=0.43]{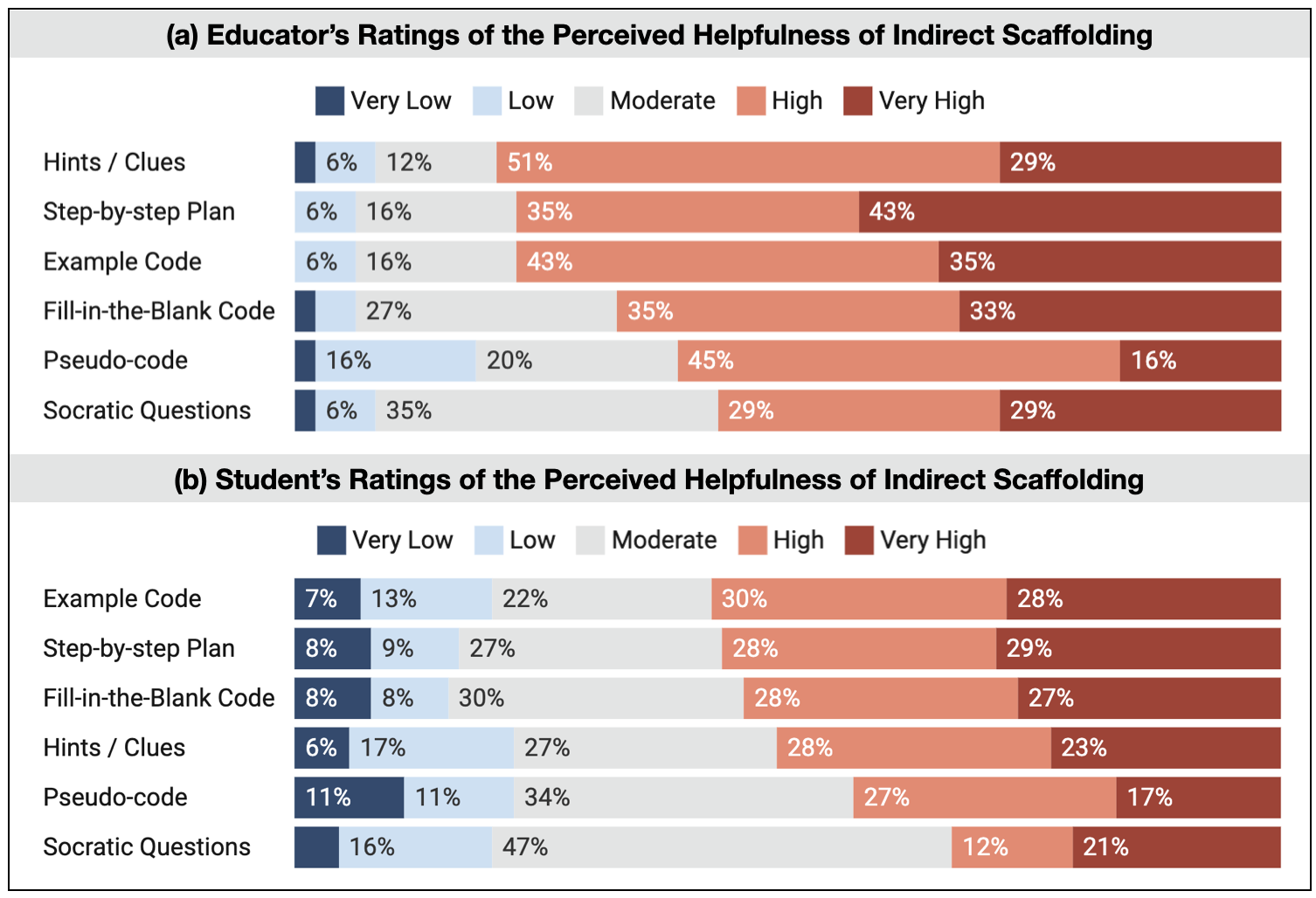}
\caption{(a) Educators' and (b) students' perceived helpfulness of indirect scaffolding. Items are ordered by the sum of High and Very High responses.}
\label{survey2}
\end{figure}

\subsubsection{Perceived Helpfulness of Indirect Scaffolding}

Figure~\ref{survey2}(a) summarizes educators’ ratings of how helpful different indirect scaffolds would be. Overall, educators rated \emph{Hints/clues}, \emph{Step-by-step plans}, and \emph{Example code} most positively, suggesting that educators value actionable support while still requiring students to complete the target solution themselves. Also, \emph{Fill-in-the-blank code} was viewed favorably. In contrast, \emph{Pseudo-code} and \emph{Socratic questions} received fewer positive evaluations, suggesting that some educators may view them as either too abstract to be practically helpful or too closely tied to solution structure. 

Figure~\ref{survey2}(b) shows a different pattern among students. Students rated \emph{Example code} most positively overall, followed closely by \emph{Step-by-step plan} and \emph{Fill-in-the-blank code}. \emph{Hints/clues} was evaluated more moderately. Similar to educators, \emph{Pseudo-code} and \emph{Socratic questions} were the least favored scaffold for students, indicating that many students do not experience purely question-driven guidance as helpful in the moment.


\subsubsection{Max Level of Scaffolding Boundary}

Figure~\ref{summary}(c-2) shows the result about the maximum level of scaffolding that should be permitted in a learning-oriented assistant, educators’ preferences concentrated around mid-level supports rather than full solutions. The most frequently selected upper bounds were \emph{Up to a step-by-step plan} and \emph{Up to example code}, followed by \emph{Up to fill-in-the-blank code}. Only a small minority endorsed allowing \emph{Up to direct runnable code} as the maximum, while several educators preferred \emph{No AI assistance} at all. Overall, they tended to endorse “bounded help” that supports decomposition and understanding, but stops short of runnable solutions that could short-circuit students’ reasoning or encourage over-reliance.

Students selected higher ceilings on average, though their most common choice still focused on structured guidance rather than complete solutions. The modal preference was \emph{Up to a step-by-step plan}, followed by \emph{Up to example code}. However, compared with educators, students were more willing to allow extremes at both ends, with more selecting \emph{No AI assistance}, \emph{Only hints/clues}, as well as \emph{Up to direct runnable code}. 

\subsection{Control and Responsibility}

As shown in Figure~\ref{summary}(d), both educators and students most favored a shared-control arrangement in which \emph{Instructors set a maximum level of assistance, and students choose within that range}. At the same time, educators more often endorsed \emph{Student chooses freely} than \emph{System suggests; student confirms}, and very few wanted the \emph{System controls all}. This pattern suggests that while educators want to enforce scaffolding, they remain cautious about delegating pedagogical authority to the AI system itself. Students, by contrast, showed a more distributed set of preferences: many supported \emph{Student chooses freely}, while a substantial group preferred \emph{Instructor controls all} or \emph{System suggests; student confirms}. Same as educator, very few students preferred \emph{System controls all}, as it may reduce students’ autonomy and transparency over how assistance levels are determined.

\section{Discussion and Design Implications}

In this section, we answer our research questions and derive design guidelines for AI coding assistants in programming education.

\textbf{Policy and Perceptions (RQ1).}
Across both groups, AI was generally viewed as beneficial for learning, and "allowed with conditions" was the most common policy preference, suggesting that stakeholders accept AI assistance when it is paired with clear boundaries. One implication is to translate policy into enforceable interaction constraints: instructors should be able to specify what the assistant may do according to different learning contexts, and the assistant should consistently reflect these rules in its outputs. Making constraints visible reduces ambiguity and helps align AI use with course goals, rather than relying on students to infer expectations or self-police.

\textbf{Learner Input and Context (RQ2).}
Although educators and students prioritized these choices differently, both groups showed strong support for each side of the trade-off—manual vs.\ automatic context provision and free-form vs.\ structured input. This pattern likely reflects variation in help-seeking situations and user expertise. When students have a clear question and sufficient experience, they tend to prefer free-form input so they can specify details in their own words. In contrast, when students are unsure how to begin—such as when they cannot identify what information is missing or how to describe their issue—they expect clearer initial guidance and benefit from the system automatically gathering relevant context to make the assistance actionable \cite{ma2025scaffolding,scholz2025partnering}. This suggests that an effective assistant should support switching between modes: offering lightweight structure for novices or early use, while allowing that structure to fade as students become more confident and develop effective help-seeking routines.

\textbf{AI Responses and Scaffolding (RQ3).}
Our results show educators tended to prefer indirect scaffolding, while students more often wanted direct, actionable help when stuck. This suggests that a single response style is unlikely to satisfy both instructional goals and user expectations. One implication is to implement progressive, learning-oriented disclosure in AI coding assistance. Instead of providing a single, fixed response style, systems can begin with indirect scaffolding and permit controlled escalation toward code-oriented guidance only when students remain blocked  \cite{kazemitabaar2024codeaid,ma2025scaffolding}. When revealing code becomes necessary, the interface can incorporate accountability mechanisms—such as requiring brief explanations or comprehension checks on critical code segments—to preserve cognitive engagement and reduce answer-copying risks. In addition, the assistant should explicitly distinguish code examples from direct runnable solutions, and condition the availability of each on instructional context (e.g., practice versus graded assessments), thereby balancing actionability with pedagogical constraints \cite{kazemitabaar2024codeaid}.

\textbf{Control and Responsibility (RQ4).}
Both educators and students most favored shared control: instructors set a maximum assistance level, and students choose within that range. This model preserves instructional authority (course-aligned ceilings) while allowing students flexibility to adjust help based on their moment-to-moment needs and individual preferences. Prior work further suggests that students’ desired degree of control can vary with individual factors such as self-efficacy and metacognitive skills, with more confident students often seeking greater autonomy \cite{wu2025learner,brusilovsky2024ai}. Taken together, these findings motivate adaptive shared-control designs that keep instructor-defined boundaries stable while supporting student choice and transparency within those bounds.

\section{Conclusion}
This paper discussed how AI coding assistants should be designed during programming practice in classrooms. We compared educator and student preferences from four design considerations. Educators generally favored indirect scaffolding and course-aligned constraints, whereas students more often preferred direct, actionable help. Despite this tension, both groups converged on a shared-control model in which instructors set a maximum help level and students choose within bounds. We hope the results from our study could help guide the future development of AI-powered coding assistants.

%
%


%
%
%
%

\bibliographystyle{splncs04}
\bibliography{bib}

\end{document}